\documentclass{article}
\pdfoutput=1
\usepackage{epsfig}

\newcommand{\ie}{{\it i.e.\ }}
\newcommand{\etal}{{\it et al.\ }}

\begin{document}

\title{Protein Interaction Networks - More than Mere Modules}

\author{Stefan Pinkert$^{1}$,  J\"org Schultz$^1$ \& J\"org
  Reichardt$^{2}$\footnote{to whom correspondence should be addressed:
    reichardt@physik.uni-wuerzburg.de } \\
 $^{1}$Department of Bioinformatics, Biocenter\\
 $^{2}$Institute of Theoretical Physics\\
  Am Hubland, University W\"urzburg}

\newpage

\maketitle

\textbf{Author Summary (193 words):}
\begin{quote}
Cellular function is widely believed to be organized in a modular fashion. On all scales and at all levels of complexity, relatively independent sub-units perform relatively independent sub-tasks of biological function. This functional modularity must be reflected in the topology of molecular networks. But how a functional module should be represented in an interaction network is an open question. In protein-interaction networks (PIN), one can identify a protein-complex as a module on a small scale, \ie modules are understood as densely linked, resp. interacting, groups of proteins, that are only sparsely interacting with the rest of the network. 

In this contribution, we show that extrapolating this concept of cohesively linked clusters of proteins as modules to the scale of the entire PIN inevitable misses important and functionally relevant structure inherent in the network. As an alternative, we introduce a novel way of decomposing a network into functional roles and show that this represents network structure and function more efficiently. This finding should have a profound impact on all module assisted methods of protein function prediction and should shed new light on how functional modules can be represented in molecular interaction networks in general.  
\end{quote}


\textbf{Abstract (302 words):}
\begin{quote}
It is widely believed that the modular organization of cellular function is reflected in a modular structure of molecular networks. A common view is that a ``module'' in a network is a cohesively linked group of nodes, densely connected internally and sparsely interacting with the rest of the network.  Many algorithms try to identify functional modules in protein-interaction networks (PIN) by searching for such cohesive groups of proteins. 

Here, we present an alternative approach independent of any prior definition of what actually constitutes a ``module''. In a self-consitent manner, proteins are grouped into ``functional roles'', if they interact in similar ways with other proteins according to their functional roles. Such grouping may well result in cohesive modules again, but only if the network structure actually supports this. 

We applied our method to the PIN from the Human Protein Reference Database and found that a representation of the network in terms of cohesive modules, at least on a global scale, does not optimally represent the network's structure because it focusses on finding independent groups of proteins. In contrast, a decomposition into functional roles is able to depict the structure much better as it also takes into account the interdependencies between roles and even allows groupings based on the absence of interactions between proteins in the same functional role, as is the case for transmembrane proteins, which could never be recognized as a cohesive group of nodes in a PIN.

When mapping experimental methods onto the groups, we identified profound differences in the coverage suggesting that our method is able to capture experimental bias in the data, too. For example yeast-two-hybrid data were highly overrepresented in one particular group. 

Thus, there is more structure in protein-interaction networks than cohesive modules alone and we believe this finding can significantly improve automated function prediction algorithms in the future
\end{quote}

{\bf Abbreviations:} PPI, protein protein interaction; GO, Gene Ontology; HPRD, Human Protein Reference Database\\


\section{Introduction}
Biological function is believed to be organized in a modular and hierarchical fashion \cite{BarabasiNetBio}. Genes make proteins, proteins from cells, cells form organs, organs form organisms, organisms form populations and populations form ecosystems. While the higher levels of this hierarchy are well understood, and the genetic code has been deciphered, the unraveling of the inner workings of the proteome poses one of the greatest challenges in the post-genomic era \cite{FuncPredReview}. The development of high-throughput experimental techniques for the delineation of protein-protein interactions as well as modern data warehousing technologies to make data available and searchable are key steps towards understanding the architecture and eventually function of the cellular network. These data now allow for searching for functional modules within these networks by computational approaches and for assigning of putative protein functions based on such data.

A recent review by Sharan \etal \cite{FuncPredReview} surveys the current methods of network based prediction methods for protein function. Proteins must interact to function. Hence, we can expect protein function to be encoded in a protein interaction network. The basic underlying assumption of all methods of automated functional annotation is that pairwise interaction is a strong indication for common function. 

Sharan \etal differentiate two basic approaches of network based function prediction: ``direct methods'', which can be seen as local methods applying a ``guilt-by-association''  principle \cite{GlobalGuilt} to immediate or second neighbors in the network, and ``module assisted'' methods which first cluster the network into modules according to some definition and then annotate proteins inside a module based on known annotations of other proteins in the module. So instead of ``guilt-by-association'', one could speak of ``kin-liability''. The latter approach to function prediction necessarily needs a concept of what is to be considered a module in a network. Most researchers consider cohesive sets of proteins which are highly
connected internally, but only sparsely with the rest of the network \cite{pmid14517352,pmid18385821,pmid17147822,Palla05,AdamcsekCFinder,BuYeast,DunnPPI,ComplexPrediction,Krognan,PereiraPPI,PrzuljPPI}. Such methods have yielded considerable success at the level of very small scale modules and in particular protein complexes.

Does the concept of a module as a group of cohesively interacting proteins also extend to larger scales? Some researchers have argued that modularity in this sense is a universal principle such that small cohesive modules combine to form larger cohesive entities in a nested hierarchy \cite{Ravasz,ClausetHierarchical}.
But is this view really adequate to describe the architecture
of protein interactions? Recently, Wang and Zhang \cite{pmid17542644} even questioned whether cohesive clusters in protein interaction networks do carry biological information at all and suggested a simple network growth model based on gene duplication which would produce the observed structural cohesiveness as ``an evolutionary byproduct without biological significance''.  We will not go as far as questioning the content of biological information in the network structure but rather argue against the model of a cohesively linked group of nodes in a network as an adequate proxy for a functional module on all scales of the network. 

Consider as first example protein
complexes. Indeed, they consist of proteins working together and
experimentally isolated together. Only the large scale analysis of
protein complexes \cite{pmid16429126, pmid11805826} revealed that they are more dynamic
than previously assumed. Many proteins can be found not only in a
single, but in a multitude of complexes. The information of proteins
connecting complexes will be lost when searching only for cohesively
interacting groups of proteins. As a second example, consider transmembrane
proteins, like receptors in signal transduction cascades. They tend to
interact with many different cytoplasmic proteins as well as with
their extracellular ligands. Still, only rarely do different
transmembrane receptors interact with each other. Thus, the functional
class of transmembrane receptors will not be identified when looking
for cohesive modules.

Here, we asked whether these features, which are not covered by algorithms searching for cohesive modules, are also present in the overall structure of the cellular network. If this would be the case, methods searching only for cohesive modules would not be able to identify them. We group proteins self-consistently into \emph{functional roles} if they interact in similar ways with other proteins according to their functional roles. Such a role may well be a cohesive module, meaning that proteins in this class predominantly interact with other proteins of this class, but it does not have to. In other words, we do not impose a structure of cohesive modules on the network in our analysis but rather find the structural representation that is best supported by the data.
Using the abstraction of a functional role, we generated an 'image graph' of the original network which depicts only the predominant interactions among classes of proteins and thus allowing a bird's eye view of the network. 

In the case of protein interaction network studied here, we found sound evidence that cohesive modules on a global scale do not adequately represent the network's global structure.  We found groups of
proteins acting as intermediates and specifically connecting other groups of proteins. Furthermore, we
even identified a group of proteins which was only sparsely connected within itself, but
with similar patterns of interaction to other proteins. Thus,
approaches searching only for cohesive modules might not be
sufficient to represent all characteristics of cellular networks. Furthermore, our findings suggest that hierarchical modularity as nested, cohesively interacting groups of proteins has to be reconsidered as a universal organizing principle. 

\section{Functional Role Decomposition and Image Graphs}
\begin{figure}[t!]
\begin{center}
\includegraphics[width=7cm]{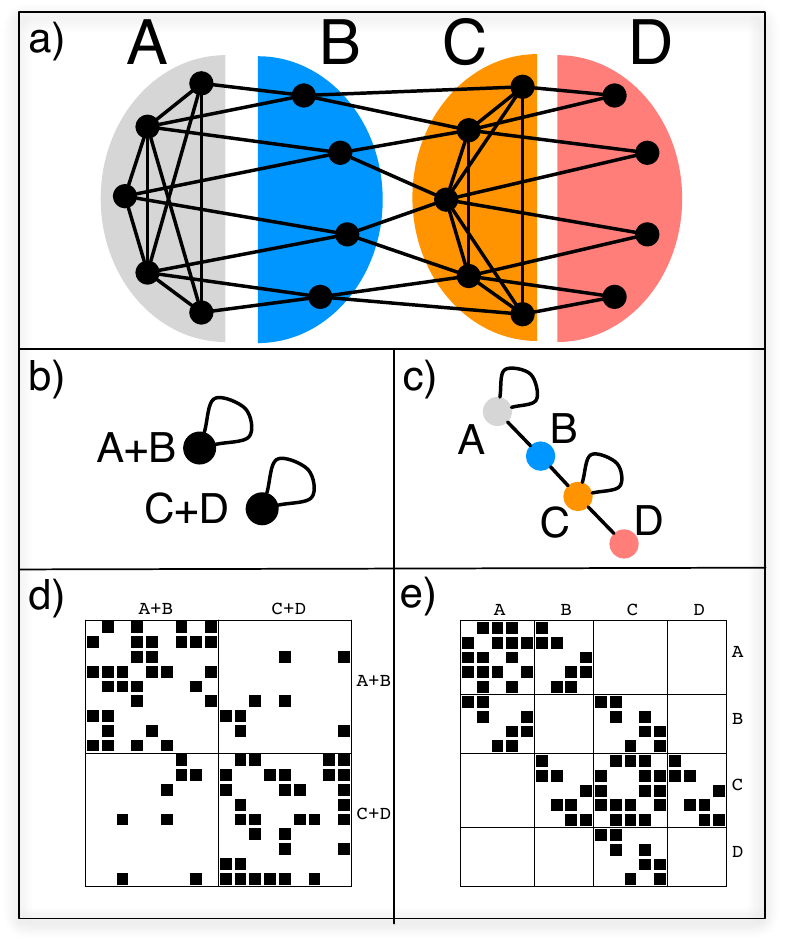}
\end{center}
\renewcommand{\baselinestretch}{1}
\caption{\textbf{An example network and possible image graphs.} {\bf a)} A simple example network of nodes of 4 different types identified by their structural position. Nodes of types A and C are densely connected among themselves. The nodes of type B have connections to both nodes of types A und B, but not among themselves, i.e. they mediate between types A and C. The nodes of type D only have connections to nodes of type C, but not among each other, i.e. they form a periphery to type C nodes. {\bf(b)} and {\bf c)} Two possible image graphs for the functional understanding of this network show the connections among groups of nodes. A typical network clustering will aggregate nodes into clusters densely connected internally but only sparsely connected to the rest, as depicted in the left image graph. This will result in grouping nodes of types A and B together and nodes of type C+D together. Because of aggregating nodes into cohesive groups any such algorithm will never recognize nodes of type C and D as different and hence miss essential part of the networkÍs structure. On the opposite, the right image graph  correctly captures the network structure of the 4 different types as the 4 different nodes in the image graph. {\bf d)} and {\bf e)} The adjacency matrices of our example network with rows and columns ordered according to the two decompositions shown above. A black square in position $(i,j)$ indicates the existence of a link connecting node $i$ with node $j$. Rows and columns are ordered such that nodes in the same group are adjacent. The internal order of the nodes in the groups is random. Each block in the matrix corresponds to a possible edge in the image graph. The left matrix shows the adjacency matrix for the output of a typical clustering algorithm which groups nodes of type A and B, as well as C and D together. Clearly we see dense blocks along the diagonal and sparse blocks on the off-diagonal of the matrix as expected. The right matrix depicts the adjacency matrix with rows and columns according to the actual types of the nodes. All empty blocks in this matrix correspond to a missing edge in the image graph and all populated blocks are represented by an edge in the image graph. We see that for this network, the image graph perfectly captures the structure of the network.}
\label{ExampleNet}
\end{figure}

In which cases does a clustering of a network into cohesive
modules not reflect its original architecture? Consider the toy
network in Figure \ref{ExampleNet} a). There are four known types of
proteins in this network. Type $A$ may represents some biological process
involving five proteins connected to four proteins of type $B$. These
are linked to another biological process $C$ which involves 
five further proteins which finally are linked to four proteins of type
$D$. Not all nodes of the same type necessarily share the same set of
neighbours. Some nodes of the same type do not have any neighbours in
common with nodes of their type or have more neighbours in common with
nodes of a different type. \enlargethispage{0.5cm}This shows that in this hypothetical example, direct methods of functional annotations may be limited in their accuracy.

\clearpage

Clustering the network into cohesive
modules cannot capture the full structure of the
network. The nodes of type B will never be recognized as a proper
cluster, because they are not connected internally at all. An attempt
to identify such groups was made by Guimera \etal
who quantified the error to such a cohesive clustering approach in a
``participation coefficient'' which is then used to differentiate groups of proteins by this participation coefficient. \cite{pmid15729348}.

The structure of the example network can, however,  be
perfectly captured by a simple image graph with 4 nodes (Fig. \ref{ExampleNet} c). The nodes in an image graph correspond to the types of nodes in the network. Nodes of type $A$ are
connected to other nodes of type $A$ and to nodes of type $B$. Nodes
of type $B$ have connections to nodes of types $A$ and $C$ and so
forth. The concept of defining types of nodes by their relation to
other types of nodes is known as ``regular equivalence'' in the social sciences
\cite{WhiteReitz,LorrainWhite}. Structure recognition in networks can then be
seen as finding the best fitting image graph for a network.
In this context, clustering into functional modules means
representing the network by an image graph
consisting of isolated, self-linking nodes. Once an assignment of
nodes into classes is obtained, the rows and columns of
the incidence matrix can be reordered such that rows and columns
corresponding to nodes in the same class are adjacent (Fig. \ref{ExampleNet} d and e). Since the rows and columns are not ordered
within a certain class, this leads to a characteristic structure with
dense blocks in the adjacency matrix corresponding to the links in the
image graph and sparse or zero blocks corresponding to the links
absent in the image graph. Structure recognition in networks is
therefore also called ``block modelling'' and together with the
concepts of structural and regular equivalence has a long history in
the social sciences \cite{DoreianBook,WassermanFaust}. In our further discussion, we
will denote image graphs that consist only of isolated, self-linked
nodes as in Figure \ref{ExampleNet} b), ``diagonal image graphs'' due
to the block structure along the diagonal in the adjacency matrix that
they induce. Accordingly, we will call all other image graphs
``non-diagonal image graphs''.


\subsection{Calculation}
But how do we find the best fitting image graph? The problem amounts
essentially to aligning a small graph with $q$ nodes to a large
network with $N$ nodes. This involves finding an image graph \emph{and} a mapping $\tau$ of the $N$ nodes of the network to the $q$
types of nodes such that the mismatch between network and image graph
is minimal. Suppose we were given the $q\times q$ adjacency matrix
$B_{rs}$ of our image graph together with the $N\times N$ adjacency
matrix $A_{ij}$ of our network . Let $\tau$ be the mapping of the $N$
nodes to the $q$ different types, such that $\tau_i\in \{1,..,q\}$ for
all $i\in \{1,..,N\}$. To optimize the
mapping $\tau$  we minimize the following error function:
 \begin{eqnarray}E(\tau,B) & = &\frac{1}{M}\sum_{i\neq j}^N(A_{ij}-B_{\tau_i\tau_j})(w_{ij}-p_{ij})\label{Error}\\  
 & = &\underbrace{\frac{1}{M}\sum_{i\neq j}^N (w_{ij}-p_{ij})A_{ij}}_{\mathcal{Q}_{\mbox{\tiny max}}< 1}-\underbrace{\frac{1}{M}\sum_{i\neq j}^N(w_{ij}-p_{ij}) B_{\tau_i\tau_j}}_{\mathcal{Q}(\tau,B)\leq\mathcal{Q}_{\mbox{\tiny max}}}.\label{ErrorFit}
 \end{eqnarray}
in which $A_{ij}$ is the $\{0,1\}$ adjacency matrix of the network
under study.  $w_{ij}$ denotes the weight
given to an edge between nodes $i$ and $j$. If an edge is absent in
the network, $w_{ij}$ is naturally zero. As before $B_{\tau_i\tau_j}$
is the image graph and $p_{ij}$ is a penalty term discussed below. The
normalization constant $M=\sum_{i\neq j}w_{ij}$ is used to bound the
error by one. This error function gives a weight proportional to
$(w_{ij}-p_{ij})$ to errors made on fitting the edges in the network
and a weight of $p_{ij}$ to errors made on fitting the absent edges in
the network. The penalty term
$p_{ij}$ is chosen such that the total error weight on all edges in
the network is equal to the total error weight on all absent edges in
the network: 
\begin{equation}
\sum_{i\neq j}^NA_{ij}(w_{ij}-p_{ij})= \sum_{i\neq j}^N(1-A_{ij})p_{ij}.
\label{PenaltyDef}
\end{equation}
This can be easily achieved by setting $p_{ij}=(\sum_{k\neq i} w_{ik}
\sum_{l\neq j} w_{lj})/\sum_{k\neq l} w_{kl}$. The first term of
equation (\ref{ErrorFit}) neither depends on the
mapping of nodes to types $\tau$ nor on the image graph $B_{rs}$. It
can be interpreted as the maximum value of a quality function
$\mathcal{Q}$ measuring the fit of the image graph to the network
which would be obtained for a perfect fit, \ie
$B_{\tau_i\tau_j}=A_{ij}$ for all $(i,j)$. The second term then
corresponds to the quality of the actual fit for the given image graph
and mapping. The error is simply the difference between the best and
any sub-optimal fit and minimizing $E$ and maximizing $\mathcal{Q}$
are equivalent.  

If we assume a diagonal image graph $B_{rs}=\delta_{rs}$ we recover in
$\mathcal{Q}$ of equation (\ref{ErrorFit}) a popular quality function
for graph clustering known as Newman modularity \cite{Girvan03,pmid15729348,pmid17542644}. We can
hence directly compare the fit of different given image graphs to one
network by the maximum score $\mathcal{Q}$ than can be obtained by
optimizing the mapping $\tau$ of nodes in the network to the classes
represented as nodes in that image graph.  
The overall optimal image
graph with a given number of nodes $q$ and the optimal assignment $\tau$ into the $q$ classes can be found directly by searching for the assignment $\tau$ which maximizes \cite{ReichardtWhite,ReichardtLNP}
\begin{equation}
\mathcal{Q}^*(\tau)=\frac{1}{2M}\sum_{r,s}^q ||\sum_{i\neq j}^N (w_{ij}-p_{ij}) \delta_{\tau_ir} \delta_{\tau_js}||.
\label{Qstar}
\end{equation}
The image graph which allows the highest value of $\mathcal{Q}$ among all
possible image graphs with this number of classes can be read off from the assignment $\tau$ that maximizes (\ref{Qstar}). It must be such that $B_{rs}=1$, if the argument in the absolute
value in (\ref{Qstar}) is strictly positive, and zero otherwise. One can view $B_{rs}$ as a lossy compression of the original network, in contrast to recently introduced lossless network compression methods for biological analysis \cite{PowerGraph}. Since most of the currently available data on protein interaction is noisy and incomplete, we find a lossy compression most adequate for the analysis of the large scale structure of the network.    

\section{Results}
\subsection{Network analysis}
\begin{figure}[h]
\hspace{-0.8cm}\includegraphics[width=8.7cm]{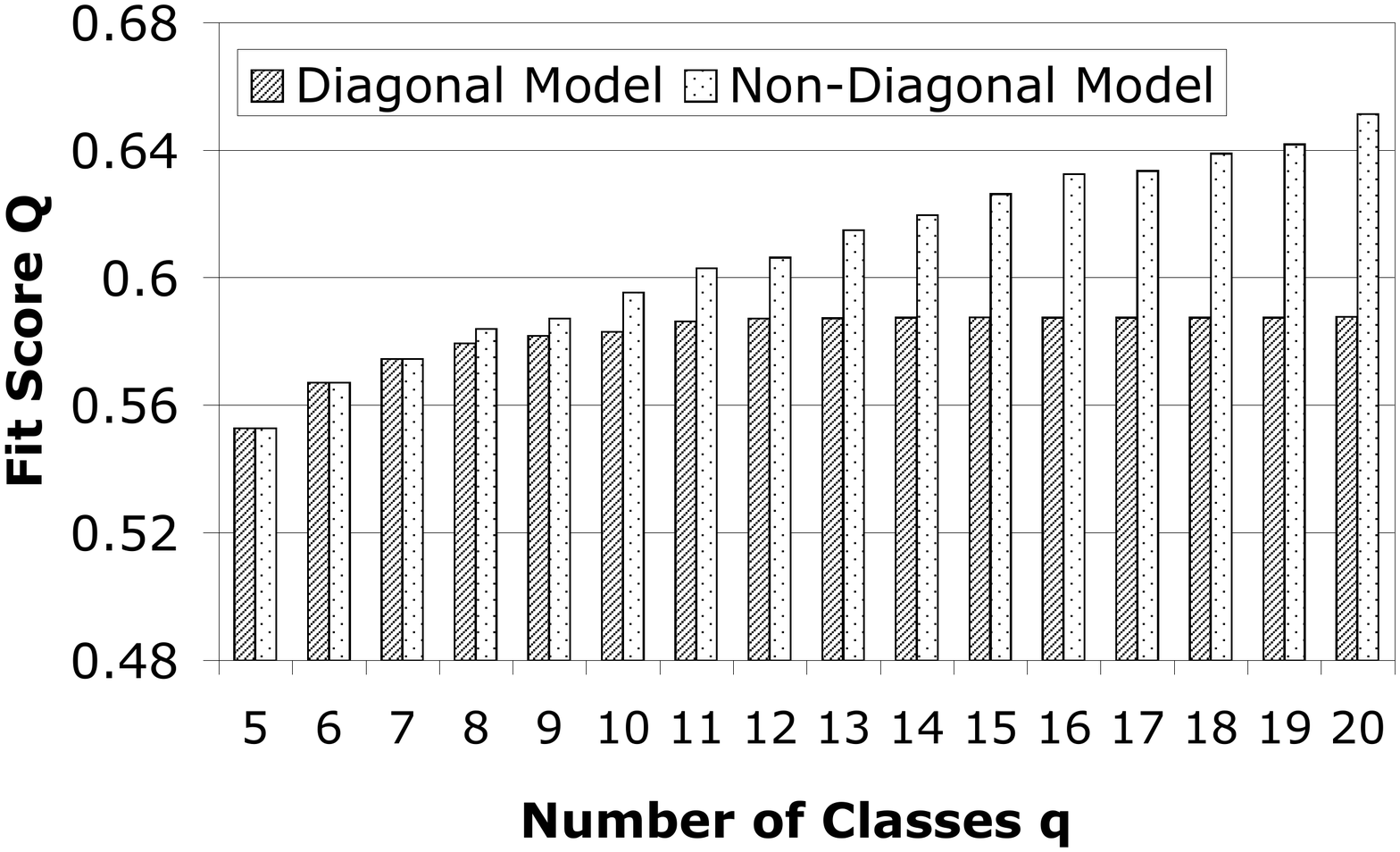}
\includegraphics[width=8.7cm]{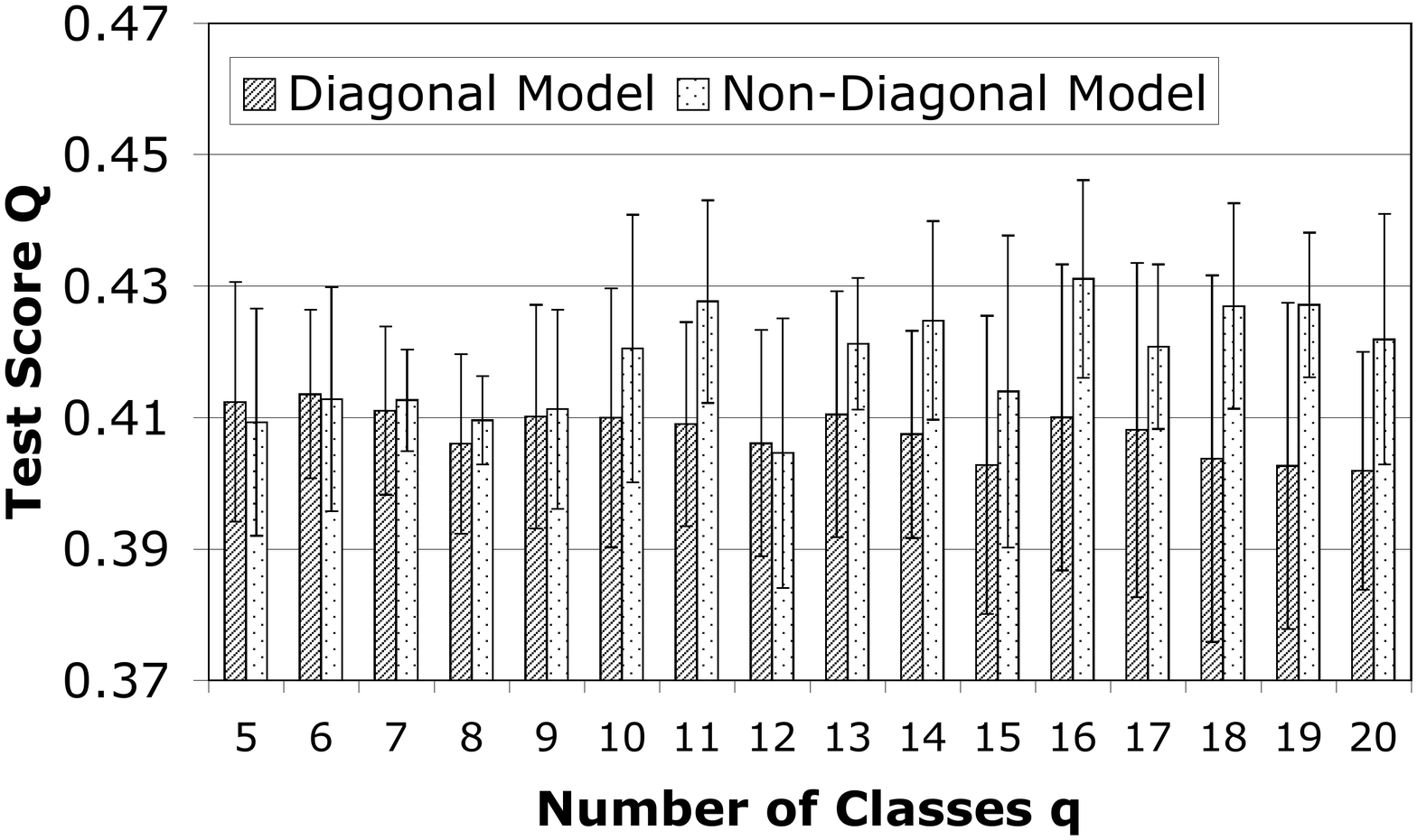}
\renewcommand{\baselinestretch}{1}
\caption{\textbf{Fit scores and generalization error.  Left: }Comparison of highest fit scores $Q$ and $\mathcal{Q}^*$ for the full dataset with 32,331 interactions. Clustering methods aggregating nodes into cohesive groups (diagonal image graphs) cannot improve the score beyond a certain limit, while non-diagonal image graphs are able to capture more and more structure as the image graph gets larger and larger. {\bf Right: }After removing 1000 links from the data as test-set, we optimized the assignment of nodes into classes according to (\ref{ErrorFit}) using only the remaining links and keeping the image graphs fixed to those found in the runs that lead to the figure on the left. With the assignment of nodes into classes for this training set of links, we computed the score on the test set of links. The figure shows average and standard deviation over 100 repetitions of this experiment.
}
\label{FitScores}
\end{figure}

Using the quality function introduced above, we analysed the HPRD protein
interaction network containing 8,500 nodes. We considered the entire
network and optimised $\mathcal{Q}^*$ from (\ref{Qstar}) - thus
finding optimal image graphs and assignments of nodes into
classes. As expected, with increasing number of classes $q$, the fit between the
actual network and the image graphs becomes better
(Fig. \ref{FitScores}, left panel). Restricting the image graphs to a diagonal
form $B_{rs}=\delta_{rs}$ also limited the fit score. The maximum fit
score was equal to $\mathcal{Q}_{\mbox{\tiny max}}=0.98$. Therefore,
even with a very small number of classes, already $2/3$ of the link
structure in the network was captured. The maximum of $\mathcal{Q}$
for a diagonal image graph was reached at $q=11$ and further addition of classes did not increase
this value any more. For $q<8$ the fit scores for diagonal and
non-diagonal image graphs were equal because for less than 8 classes
the best image graphs were in fact diagonal. Only beyond this point, the
additional degrees of freedom of the non-diagonal image graphs allowed
better fit scores. 

The question now is, whether these additional degrees
of freedom in the image graph actually convey information or only led to overfitting. We
therefore divided the $32,331$ links of the network into a test- and a
training-set of $1,000$ and $31,331$ links, respectively. Using the
optimal image graphs obtained on the full data set and diagonal image
graphs for comparison, we optimized $\mathcal{Q}$ from
(\ref{ErrorFit}) on the training-set of links and with the resulting mapping of
nodes into classes calculated the fit score $\mathcal{Q}$ on the test-set. The fit score on the training-set of links (data not shown) was close to the full data set. We fixed the non-diagonal image graphs because the comparison is made to diagonal image graphs which were unaltered, too. 

Both, diagonal and non-diagonal image graphs, showed overfitting to some
extent. The score on the test set is lower than on the training set (Fig.\ref{FitScores}, right panel).  However, for more than 8 classes, the non-diagonal
image graphs not only allowed a better fit as discussed, but also scored better on
the test-set, \ie the increased fit value also generalized! The
non-diagonal image graphs do contain more information about the
network than the diagonal image graphs. 

It has to be considered that
using a test-set containing $3.2\%$ of all links was a drastic
disturbance of the system. If we assigned nodes into $q=8$ equal sized
classes, we expect approximately $2/(q(q+1))\approx 3\%$ of all links in one block. So
above this point, the test set we removed was more than the typical
number of links in a block. Also, consider the average degree of
$\langle k\rangle\approx 8$ interactions per protein in the
network. Removing a single link means removing on average $1/8$ of the
neighbourhood of the nodes connected by this edge. For the 1,000 edges
in the test-set, this could have happened to 2,000 nodes and thus to
almost one quarter of all nodes. This explains the large 
fluctuations and may also explain that for $q=12$ the non-diagonal
image graph cannot outperform the diagonal one.

\begin{figure}[t]
\includegraphics[height=6.5cm]{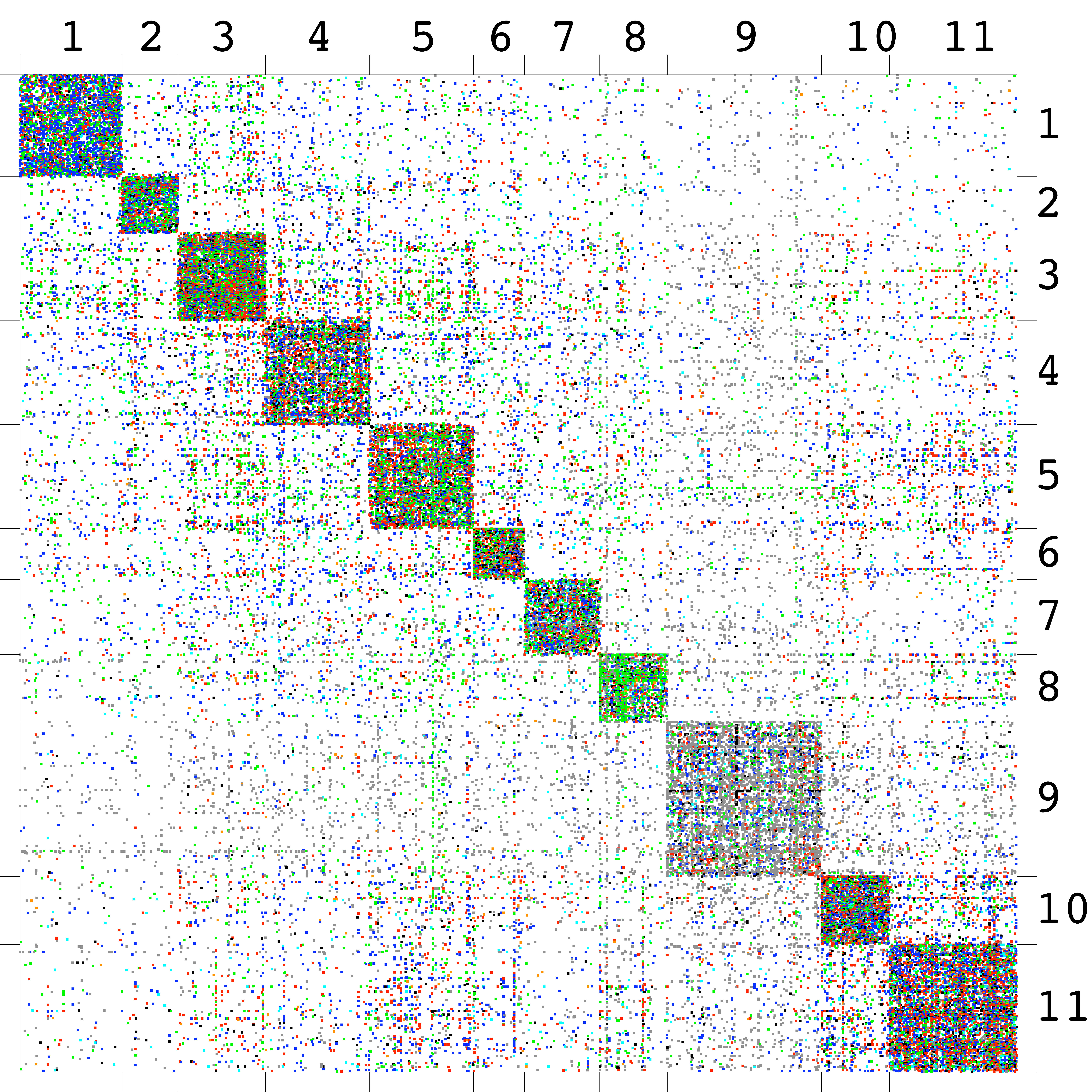}
\includegraphics[height=6.5cm]{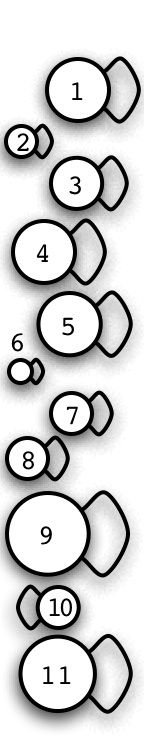}
\includegraphics[height=6.5cm]{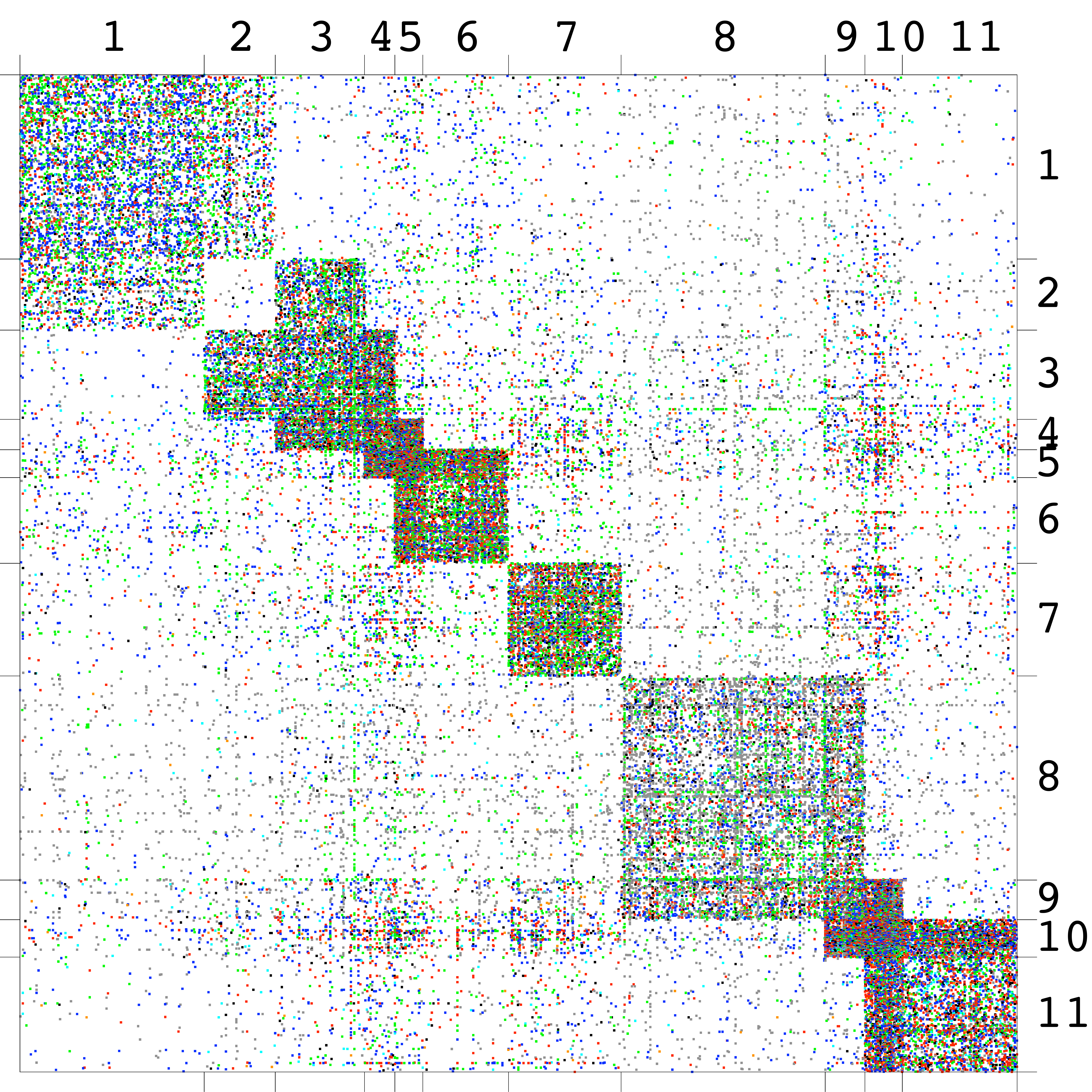}
\includegraphics[height=6.5cm]{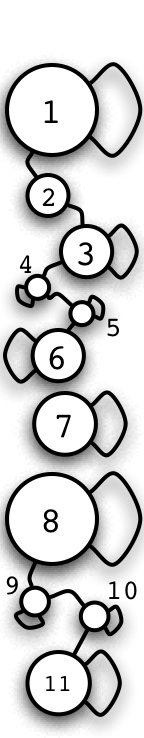}
\renewcommand{\baselinestretch}{1}
\caption{\textbf{Comparison of block assignment.} For $11$ classes, we show the adjacency matrix of the HPRD protein interaction network with rows and column ordered to show diagonal and non-diagnal block structure plus the corresponding image graphs for diagonal block models and non-diagonal block models. Note how the non-diagonal models allow to capture overlap between cohesive blocks but also to detect groups of nodes which are non-cohesive but have similar connection patterns to other classes of proteins. The color of the links codes the experiment type: Y2H: grey, in-vitro: blue, in-vitro+Y2H: turquoise, in-vivo: green, in-vivo+Y2H:orange, in-vivo+in-vitro: red, in-vivo+in-vitro+Y2H:black.
}
\label{q11Matrices}
\end{figure}

Figure \ref{q11Matrices} shows two representations of the adjacency matrix of the PIN. On the left hand side, rows and columns are ordered according to the assignment of nodes in classes when fitting a diagonal image graph, \ie when searching for cohesive modules. On the right hand side, the rows and columns are ordered according to the assignment of nodes into classes with the highest scoring non-diagonal image graphs. In both cases we allowed for 11
classes. We have chosen this number of classes because the diagonal
models did not achieve larger scores when allowing more classes. The
non-diagonal image graphs led to a different assignment of
nodes into classes with higher score but further increase of the
number of classes did not lead to significant improvement in the
generalization error (Fig. \ref{FitScores}, right panel). Note the
similarities and differences in the matrix when ordered after fitting
a diagonal image graph and after fitting a non-diagonal image
graph. 

The non-diagonal models also allowed capturing groups of
proteins that mediate between cohesive clusters such as group $2$ or that form a cohesive overlap between
cohesive clusters, such as groups $4$ and $5$ or $9$ and $10$.  

\subsection{Biological interpretation}
When comparing the cohesive module to the functional role model
(Fig. \ref{q11Matrices}) the most distinguishing feature was the
existence of connections between sets of proteins in the
latter. Groups of proteins existed, which all performed the same
``functional role'' of connecting two other groups of proteins. Thus,
a separation of the cellular network into cohesive modules omits
distinct characteristics of the network. In the functional role model,
groups were connected to other groups by a distinct set of additional
proteins. These 'connector groups' may well be themselves cohesive,
but do not have to. This was illustrated by class 2, where most of the
proteins are not interacting with other proteins in the class, but
with those of groups 1 and 3.

To evaluate the biological significance of this
result, we performed a Gene Ontology enrichment analysis for all
clusters. Class 2 was significantly ($E < 10^{-27}$) enriched in proteins
annotated as belonging to the membrane and plasma membrane
compartment. Indeed, this class contained many transmembrane
proteins like for example Cadherin. These proteins typically do not
interact with many other transmembrane proteins, but with their
extracellular binding partners and, in the case of transmembrane
receptors, with cytoplasmic signal transmitters. Indeed we found that group 1, highly interacting with proteins of class 2, mainly
consisted of proteins localised in the extracellular region ($E =
2.54E^{-168}$). Furthermore, group 3 also strongly interacting with
proteins of class 2, was enriched in proteins associated with the
plasma membrane ($E = 2.84E^{-28}$) and involved in signal transduction ($E
= 2.72E^{-20}$). Thus, the transmembrane proteins of class 2 are the
perfect biological implementation of proteins not interacting with
each other, but with proteins of defined other classes (nodes of type
B in figure \ref{ExampleNet} a). A complete GO annotation of all
clusters of classifications into $q=5$ to $q=11$ classes is given in
our supporting material at
\texttt{http://domains.bioapps.biozentrum.uni-wuerzburg.de/ppi}.

\begin{figure}[t]
\includegraphics[height=5cm]{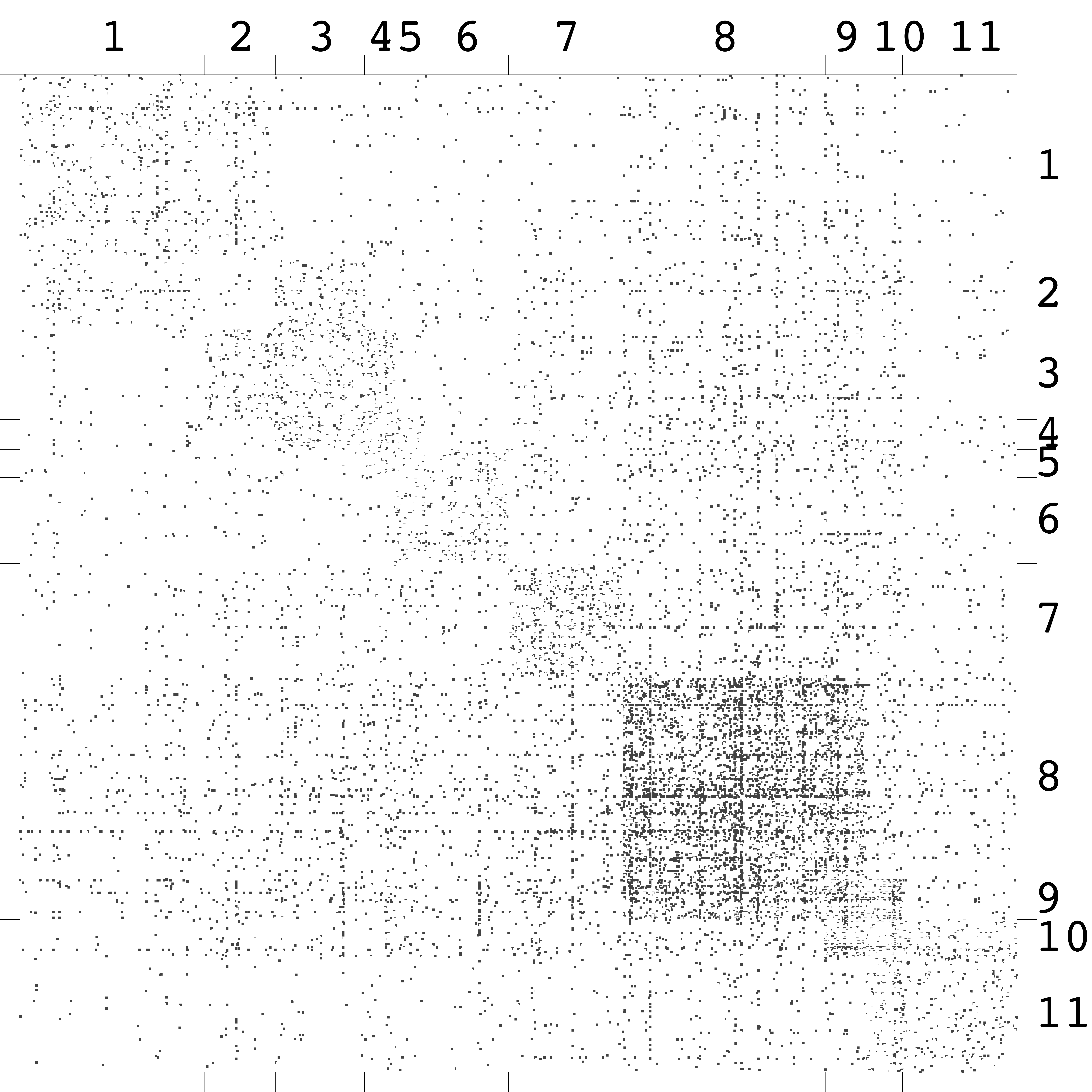}
\includegraphics[height=5cm]{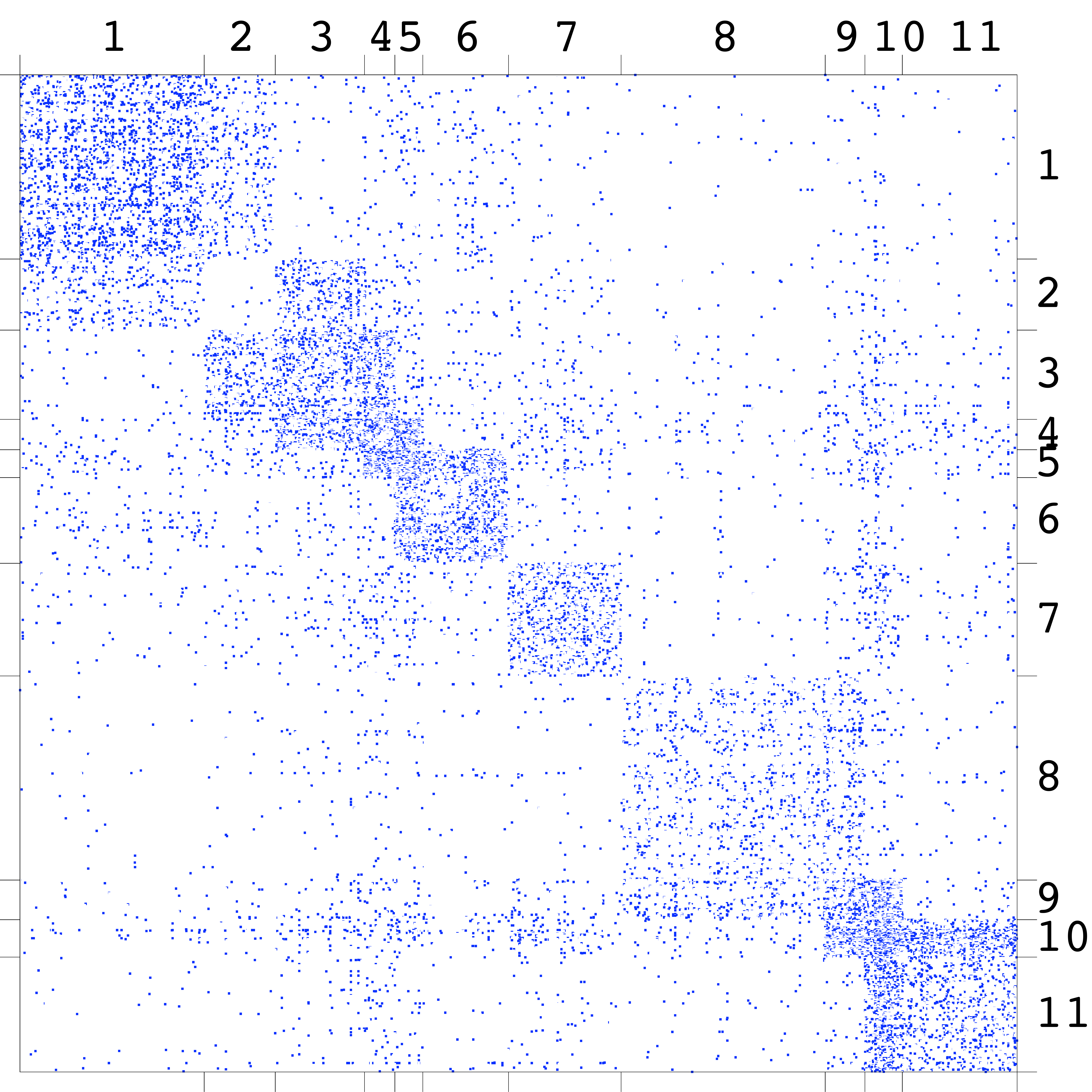}
\includegraphics[height=5cm]{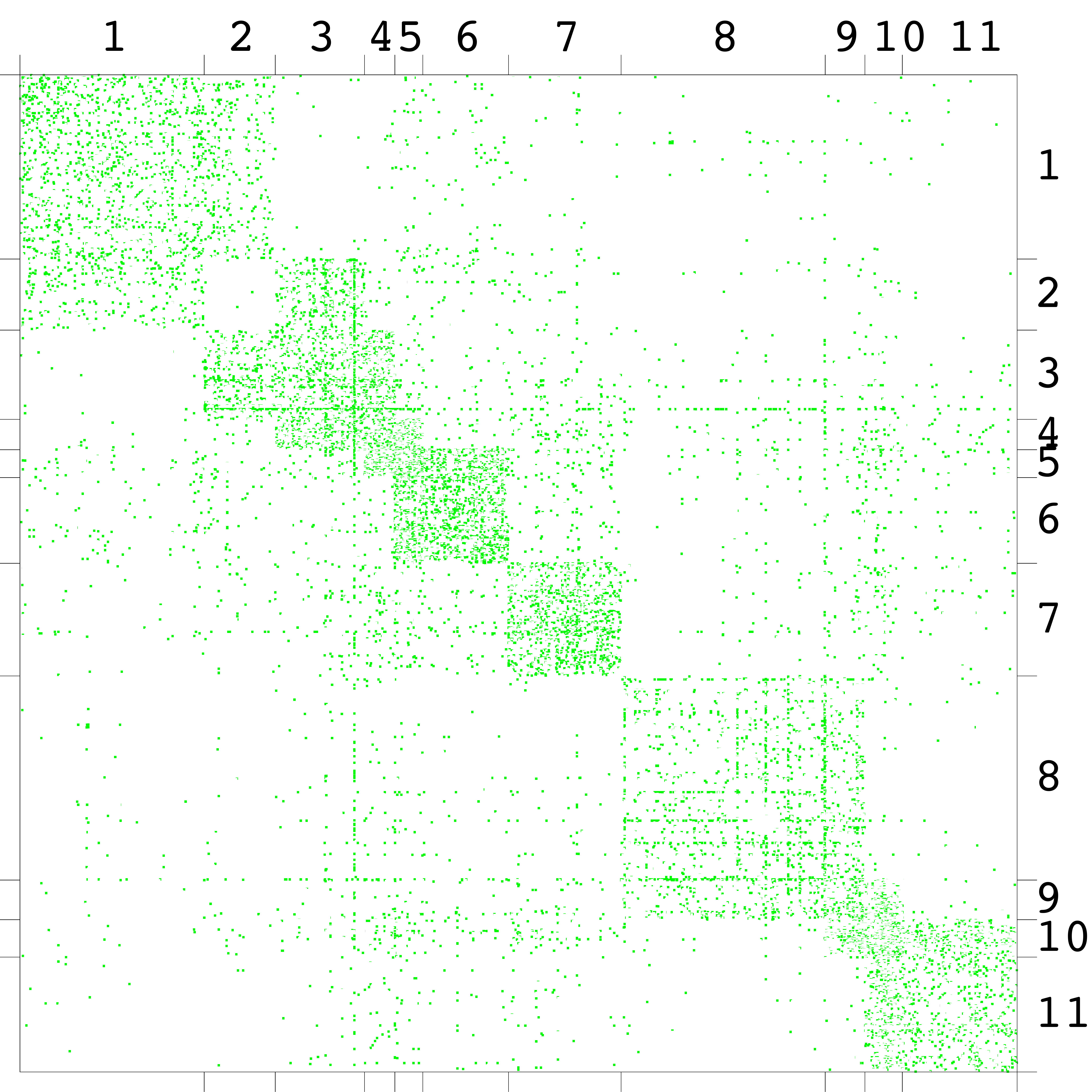}
\renewcommand{\baselinestretch}{1}
\caption{\textbf{Comparison of block assignment.} The same assignment of nodes into classes as used in Figure \ref{q11Matrices} for but 3 different types of interactions separately. {\bf Left: } Interactions reported only for yeast-2-hybrid experiments (gray). {\bf Middle: } Interactions reported only in in-vitro experiments (blue). {\bf Right: }Interactions reported only in in-vivo experiments (green). While in-vitro and in-vivo data is highly correlated, the interactions found in Y2H experiments are enriched in class 8. 
}
\label{DiffLinkMatrices}
\end{figure}

In the previous analyses, we considered all data from HPRD, as they
are manually curated and therefore of a high quality. To unravel a
possible bias btween different experimental methods, we plotted the
data for three different experimental approaches separately. The
ordering of rows and columns, \ie the assignment of proteins into
functional roles, was kept from figure \ref{q11Matrices}. Instead of
plotting all types of interactions on top of each other, the adjacency
matrices for interactions which are backed by in-vivo, in-vitro and
yeast-two-hybrid \cite{pmid2547163} (Y2H) experiments were shown
separatly (Fig. \ref{DiffLinkMatrices}).  The in-vitro and in-vivo
data nicely resembled the overall picture while the Y2H data did not
follow this pattern. To test how well the overall model described the
three experimental methods, we calculated the fit function
$\mathcal{Q}$ for each. Here, the assignment of nodes into functional
roles was taken from figure \ref{q11Matrices}. The fit score for the
interactions backed only by Y2H experiments was much lower than the
scores of any of the other experimental methods. Thus the Y2H
interactions cannot depict the full range of possible protein-protein
interactions. Rather, the data based on yeast two hybrid showed a
prevalence for class number 8 in figure \ref{DiffLinkMatrices}. In
this cluster nuclear proteins were significantly over-represented
($8.42E^{-10}$). In the Y2H \cite{pmid11283351} assay, the tested
proteins are fused to parts of a transcription factor. Their
interaction is measured by the transcription of a reporter
gene. Therefore, the proteins have to be within the nucleus. Thus, a
bias towards interactions of proteins which naturally reside in the
nucleus can be expected in Y2H data.

\begin{table}[h]
\begin{center}
\caption{{\bf Fitscore for different types of interactions. } Given the assignment of nodes into $q=11$ classes and the image graphs from figure \ref{q11Matrices}, we calculated the fit score $Q$ for each type of interaction seperately with equation (\ref{ErrorFit}). Compare to figure \ref{DiffLinkMatrices} which singles out those links which are only supported by Y2H, or only in-vivo or only in-vitro experiments.}
\label{tab:Table2}
\begin{tabular}{l|cc}
{\bf Experiment type}	& {\bf Diagonal Image Graph}&{\bf	Non-Diagonal Image Graph}\\
\hline yeast 2-hybrid	&	0.28 	&	0.30 	\\in vitro 	&	0.53	&	0.56	\\in vitro + yeast 2-hybrid	&	0.51	&	0.55	\\in vivo 	&	0.60	&	0.60	\\in vivo + yeast 2-hybrid	&	0.59	&	0.62	\\in vivo + in vitro	&	0.59	&	0.61	\\in vivo + in vitro + yeast 2-hybrid	&	0.64	&	0.64	\\
\hline
	\end{tabular}
	\end{center}
\end{table}

\section{Discussion}
Using a suited algorithm, any network can be separated into cohesive groups of
nodes with more internal than external connections. Accordingly, also
protein-protein interaction networks can be divided into comparably
independent units as putative functional modules \cite{pmid14517352}. Do
these modules really reflect a typical characteristic of the cellular network? Here, we
used  an alternative approach for the
clustering of protein interactions. We grouped proteins of a similar
functional role together. The functional role was defined by the
interactions with proteins of other groups. In contrast to cohesive modules, which
are more or less independent, groups which specifically linked other groups of
proteins could be identified. Thus, an interconnectivity of biological
units as in the case of shared components in protein complexes can
also be observed at the cellular level. Using a Gene Ontology based
classification of all proteins within the modules, we found that these
roles are mainly determined by cellular localisation but also
function. Although possibly not too surprising to the biologist, this
result underlines that the classes we identified by automatic
clustering do represent a biological signal.

Using HPRD as data source, a large-scale set of interactions with, on
average, eight connections per protein
could be analysed. As HPRD contains manually curated
data, their quality should be high enough to extend the results to
higher coverage. The analysis of interactions derived by different
experimental methods revealed a bias in the coverage especially for
yeast-two-hybrid data. The great difference of the protein interactions verified only by 
Y2H to the other methods  reminds us to pay attention to the careful 
weighting of quality and quantity. As large scale binary interaction analysis were
mainly based on Y2H, using high coverage data like the one from yeast
or \emph{Drosophila melanogaster} might even blur the
signal.  Another drawback was the small amount of interactions per protein, 
which is around three to four for the yeast, fly and nematode sets analysed in the study by Wang and Zhang \cite{pmid17542644}. Still, it would be interesting to compare 
networks between different organisms to
see whether there are changes in the clusters correlated for example
with the emergence of multicellularity. But, reliable results can only
be obtained when analysing data sets of comparable quality and size \cite{ReichardtLeonePRL}.

In summary our analysis showed that protein interaction networks are
more than sparsely interacting cohesive modules. Rather, groups of proteins are
connected by distinct sets of other proteins. These may be highly connected
to each other, but do not have to be. Therefore, functional roles and
corresponding image graphs might be better descriptors for the
characteristics of a protein interaction network than cohesive 
modules alone. They may help to further improve protein function prediction based on protein-interaction networks.

\section{Materials and Methods}

\subsection{PPI network.}
We used the binary PPI data from the
HPRD \cite{pmid16381900} (Version 6). HPRD protein
identifiers and experiment types used to support their connection were
extracted. The experiment types were transformed to weights according
to table \ref{tab:Table1}. The analysis was restricted to the largest
connected component containing 32,331(out of 34,367) interactions of 8,756
proteins (out of 8,919). These interactions do not include data 
inferred from protein complexes which may introduce errors and bias into
the network structure \cite{pmid17542644}.

\begin{table}
\begin{center}

\caption{{\bf Experiment type to link weight transformation. } We valued the different experiments compiled in the HPRD database differently, giving lowest weight to interactions found in yeast-2-hybrid experiments only and highest to those interactions found in vivo, in vitro and Y2H experiments. These weights are only to represent a ranking of a practitioners belief in their validity. }
\label{tab:Table1}
\begin{tabular}{lccc}\tabularnewline
			{\bf Experiment type}&{\bf Weight}&{\bf \# of interactions }&{\bf distinct proteins involved}\\ 		\hline	

			yeast 2-hybrid&1&6,580&3,727\\
			in vitro&2&7,872&4,302\\
			in vitro+yeast 2-hybrid&3&1,298&1,523\\ 
			in vivo&4&6,721&3,826\\ 
			in vivo+yeast 2-hybrid&5&824&1,119\\ 
			in vitro+in vivo&6&6,877&3,781\\ 
			in vitro+in vivo+yeast 2-hybrid&7&2,159&2,201\\ 
			\hline
	\end{tabular}
	\end{center}

\end{table}

\subsection{Clustering.}
We optimized (\ref{Qstar}) and (\ref{ErrorFit}) using Simulated Annealing \cite{Kirkpatrick}. Details about the implementation can be found in \cite{ReichardtWhite} and \cite{ReichardtPRE}, respectively. 
To obtain the left panel of figure \ref{FitScores}, for $q=5$ to $q=20$ classes, we chose the best of 10 runs, each, for both the fit of a diagonal block model as well as the detection of a non-diagonal block model.  The cooling factor for sets with more than ten
classes was changed from 0.99 to 0.999 to decrease the false positive
rate of local optima. To obtain the right panel of figure \ref{FitScores} we randomly divided the original set of links into a test-set of 1000 links and the remaining set was used as a training-set. We used the image graphs, both diagonal and non-diagonal, found in the earlier experiment to optimize the fit score on the training-set. The data shown are the fit scores of the test set, averaged over ten different partitions of the links into training- and test-set.  

\subsection{GO Term enrichment analysis.}
The HPRD identifiers and their corresponding GO identifiers  were taken from the same
HPRD dataset as the PPI network, re-formatted and saved into a file
readable by the Ontologizer \cite{pmid18511468}. For the Ontologizer the file gene\_ontology.obo created by the GO project \cite{pmid10802651} was be downloaded.


\bibliography{HPRD_24_11_08}

\end{document}